\documentclass{article}
\usepackage{amssymb}

\begin{document}

\title{Propagation of cosmic rays in the foam-like Universe}
\author{A. A. Kirillov, E.P. Savelova, P.S. Zolotarev \\
\emph{Branch of Uljanovsky State University in Dimitrovgrad, }\\
\emph{Dimitrova str 4.,} \emph{Dimitrovgrad, 433507, Russia} }
\date{}
\maketitle

\begin{abstract}
The model of a classical spacetime foam is considered, which consists of
static wormholes embedded in Minkowski spacetime. We examine the propagation
of particles in such a medium and demonstrate that a single thin ray
undergoes a specific damping in the density of particles depending on the
traversed path and the distribution of wormholes. The missing particles are
scattered around the ray. Wormholes was shown to form DM halos around
point-like sources and, therefore, the correlation predicted between the
damping and the amount of DM may be used to verify the topological nature of
Dark Matter.
\end{abstract}

\newpage

\section{Introduction}

The nature of Dark Matter (DM) represents one of the most important and yet
unsolved problems of the modern astrophysics. Indeed, while the presence of
DM has long been known \cite{Zw} and represents a well established fact
(e.g., see Refs \cite{dm, Pr} and references therein), there is no common
agreement about what DM is. In the simplest picture DM represents some
non-baryonic particles (predicted numerously by particle physics) which
should be sufficiently heavy to be cold at the moment of recombination and
those give the basis to the standard (cold dark matter) CDM models. The
latter turn out to be very successful in reproducing properties of the
Universe at very large scales (where perturbations are still on the linear
stage of the development) which led to a wide-spread optimistic believe that
non-baryonic particles provide indeed an adequate content of DM.

However the success of CDM models at very large scales is accompanied with a
failure at smaller (of the galaxies size) scales. Indeed, cold particles
which interact only by gravity should necessary form cusps ($\rho _{DM}\sim
1/r$) in centers of galaxies\footnote{%
The presence of cusps formed by the development of adiabatic perturbations
follows straightforwardly from the conservation of the circulation theorem
in the hydrodynamics. By other words the fact that the distribution of DM
should have cusps in galaxies is equivalent to the fact that DM should
represent cold non-baryonic particles.} \cite{NFW} (see also Ref. \cite{Cusp}
where the problem of cusps in CDM is discussed in more detail), while
observations definitely show the cored ($\rho _{DM}\sim const$) \cite{Core}
distribution. The only way to destroy the cusp and get the cored
distribution is to introduce some self-interaction in DM or to consider warm
DM. Both possibilities are rejected at large scales by observing $\Delta T/T$
spectrum (e.g., see Ref. \cite{Pr} and references therein). By other words
DM displays so non-trivial properties (it is warm or self-interacting in
galaxies, however it was cold at the moment of recombination and it is still
cold on larger (than galaxies) scales) that it is difficult to find
particles capable of reconciling such observations.

These facts support the constant interest to different alternatives of the
DM hypothesis which interpret the observed discrepancy between luminous and
gravitational masses as a violation of the law of gravity. Such violations
(or modifications of general relativity (GR)) have widely been discussed,
e.g., see Refs. \cite{VN,MOND}. However, it turns out to be rather difficult
to get a modification of GR which is flexible enough to reconcile all the
variety of the observed DM halos. Moreover, the weak lensing observations of
a cluster merge in Ref.\cite{DMpaper} seem to reject most of modifications
of GR in which a non-standard gravity force scales with baryonic mass.

The more viable picture of DM phenomena was suggested in Ref. \cite{K06}
(see also references therein) and developed recently in Refs. \cite%
{KT07,KS07}. It is based on the fact that on the very early (quantum) stage
the Universe should have a foam-like topological structure \cite{wheeler}.
There are no convincing theoretical arguments of why such a foamed structure
should decay upon the quantum stage - relics of the quantum stage foam might
very well survive the cosmological expansion, thus creating a certain
distribution of wormholes in the Friedman space. Moreover, the inflationary
stage in the past \cite{inf} should enormously stretch characteristic scales
of the relic foam. The foam-like structure, in turn, was shown to be
flexible enough to account for the all the variety of DM phenomena \cite%
{K06,KT07}; for parameters of the foam may arbitrary vary in space to
produce the observed variety of DM halos in galaxies (e.g., the universal
rotation curve for spirals constructed in Ref. \cite{KT06} for the foamed
Universe perfectly fits observations). Moreover, the topological origin of
DM phenomena means that the DM halos surrounding point-like sources appear
due to the scattering on topological defects and if a source radiates, such
a halo turns out to be luminous too \cite{KT07} which seems to be the only
way to explain naturally the observed absence of DM fraction in intracluster
gas clouds \cite{DMpaper}.

While the foam-like structure of the Universe is capable of providing a
quite good description of DM phenomena, it is necessary to look for some
independent tests to verify the topological nature of DM. Effects of the
spacetime foam attract the more increasing attention (e.g., see Refs. \cite%
{K99}-\cite{Foam2} and references therein). However most of effects
considered \cite{BK07}-\cite{Foam2} assume the foamy structure at extremely
small scales (which correspond to energies higher than $200$ $GeV$). DM
phenomena, however, suggest that the characteristic scale of the spacetime
foam $L$ (and respectively of wormholes) should be of the galaxy scale,
e.g., of the order of a few $Kpc$. The fact that the fundamental length
scale for the quantum dynamics of spacetime need not be equal to the Planck
length was also discussed recently in Ref. \cite{kl07}.

In the present paper we consider the propagation of cosmic rays in the
foam-like Universe. To this end we consider the model of the spacetime foam
\cite{KS07}, which consists of a static gas of wormholes embedded in the
Minkowski space. However contrary to above mentioned papers (e.g., see Ref.
\cite{Foam} where effects of cosmic ray interactions in a small-scale foam
have been considered)) we assume that the characteristic scale of such a
foam is of the order of a galaxy size. We demonstrate that the scattering on
the topological structure is described by a specific term in the Boltzmann
equation. We show that in a foamed space a single thin ray of particles
emitted undergoes a specific damping in the density of particles depending
on the traversed path and the distribution of wormholes, while the missing
particles are scattered and form a halo around the ray. Such halo however
has a very low density and is difficult to observe. It turns out that the
damping traces rather rigidly the amount of wormholes which in the foam-like
Universe form DM halos in galaxies. Thus, there should exist a rather strong
correlation between the damping and the distribution of DM in a galaxy which
presumably can be used to verify the topological nature of Dark matter.

\section{Boltzmann equation}

In the present section for the sake of simplicity we consider the flat
Minkowski space, while the generalization to the case of Friedman models is
straightforward. Basic elements of relativistic kinetic theory can be found
in standard textbooks, e.g., Ref. \cite{kin}. Let $f\left( r,p,t\right) $ be
the number of particles in the interval of the phase space $d\Gamma
=d^{3}rd^{3}p$. This function obeys the equation
\begin{equation}
\frac{\partial f}{\partial t}+\dot{r}\frac{\partial f}{\partial r}+\dot{p}%
\frac{\partial f}{\partial p}=C[f]+\alpha \left( r,p,t\right) -\left\vert
v\right\vert \int \beta (\Gamma ,\Gamma ^{\prime })f(\Gamma ^{\prime
})d\Gamma ^{\prime }  \label{be}
\end{equation}%
where $C[f]$ stands for collisions between particles, $\alpha \left(
r,p,t\right) $ stands for the rate of emission of particles in the phase
volume $d\Gamma $, and $\beta (\Gamma ,\Gamma ^{\prime })$ describes the
scattering on wormholes. For the sake of convenience we also distinguished
the multiplier $\left\vert v\right\vert =p/m$. Our aim is to find an
explicit expression for $\beta (\Gamma ,\Gamma ^{\prime })$.

We consider first a single wormhole, which represents a couple of conjugated
spheres $S_{\pm }$ of the radius $a$ and with a distance $d=\left\vert \vec{R%
}_{+}-\vec{R}_{-}\right\vert $ between centers of spheres. The interior of
the spheres is removed and surfaces are glued together. The gluing procedure
defines the two type of wormholes passable (traversable) and impassable. The
impassable wormhole appears when before gluing we turn out one of surfaces $%
S_{\pm }$. The impassable wormhole works merely as a couple of independent
spherical mirrors (absolute mirrors, since they reflect gravitons as well).
The passable wormhole works like a couple conjugated mirrors, so that while
an incident particle falls on one mirror the reflected particle comes from
the conjugated mirror.

Consider an arbitrary point $\vec{r}$ on the sphere $S_{-}$, i.e., $\vec{r}%
\in S_{-}$ and therefore $\xi _{-}^{2}=$ $\left( \vec{r}-\vec{R}_{-}\right)
^{2}$ $=$ $a^{2}$. The gluing procedure transforms this point into a
conjugated point $\vec{r}^{\prime }\in S_{+}$ which has the form $\vec{r}%
^{\prime }=\vec{R}_{+}+\vec{\xi}_{+}$ where $\vec{\xi}_{+}$ relates to $\vec{%
\xi}_{-}$ by some rotation $\xi _{+}^{\alpha }=U_{\beta }^{\alpha }\xi
_{-}^{\beta }$. Then for the traversable wormhole we find
\[
\int \beta (\Gamma ,\Gamma ^{\prime })f(\Gamma ^{\prime })d\Gamma ^{\prime
}=\left( f-f_{+}^{\prime }\right) \delta \left( \xi _{+}-a\right) +\left(
f-f_{-}^{\prime }\right) \delta \left( \xi _{-}-a\right) ,
\]%
where we used the notations $\vec{\xi}_{\pm }=\vec{r}-\vec{R}_{\pm }$, $%
f_{\pm }^{\prime }=f\left( r_{\pm },p_{\pm },t\right) $,
\[
\vec{r}_{\pm }=\vec{R}_{\mp }+U^{\mp 1}\vec{\xi}_{\pm },
\]%
\begin{equation}
\vec{p}_{\pm }=U^{\mp 1}\left( \vec{p}-2\left( pn_{\pm }\right) \vec{n}_{\pm
}\right) ,  \label{trf}
\end{equation}%
and $\vec{n}_{\pm }=\vec{\xi}_{\pm }/a$. This defines the scattering matrix $%
\beta (\Gamma ,\Gamma ^{\prime })$ for a single wormhole in the form $\beta
(\Gamma ,\Gamma ^{\prime })$ $=$ $\beta _{+}(\Gamma ,\Gamma ^{\prime })$ $+$
$\beta _{-}(\Gamma ,\Gamma ^{\prime })$ where%
\begin{equation}
\beta _{\pm }(\Gamma ,\Gamma ^{\prime })=\delta \left( \xi _{\pm }-a\right)
\left[ \delta \left( r-r^{\prime }\right) \delta \left( p-p^{\prime }\right)
-\delta \left( r_{\pm }-r^{\prime }\right) \delta \left( p_{\pm }-p^{\prime
}\right) \right] .  \label{tw}
\end{equation}%
In the case of a spherical mirror (i.e., of the impassable wormhole) \ this
expression reduces to the more simple form%
\begin{equation}
\beta (\Gamma ,\Gamma ^{\prime })=\delta \left( |r-R|-a\right) \delta \left(
r-r^{\prime }\right) \left[ \delta \left( p-p^{\prime }\right) -\delta
\left( p_{1}-p^{\prime }\right) \right] ,  \label{iw}
\end{equation}%
where $\vec{p}_{1}=\vec{p}-2\left( pn\right) \vec{n}$.

Let $F\left( R_{\pm },a,U\right) $ be the density of wormholes with
parameters $R_{-}$, $R_{+}$, $U$ and $a$, i.e.,
\begin{equation}
F\left( R_{\pm },a,U\right) =\sum\limits_{n}\delta \left( \vec{R}_{-}-\vec{R}%
_{-}^{n}\right) \delta \left( \vec{R}_{+}-\vec{R}_{+}^{n}\right) \delta
\left( a-a_{n}\right) \delta \left( U-U_{n}\right) .  \label{F}
\end{equation}%
Then the total scattering matrix is described by
\begin{equation}
\beta _{\pm }^{tot}(\Gamma ,\Gamma ^{\prime })=\int \beta _{\pm }(\Gamma
,\Gamma ^{\prime })F\left( R_{\pm },a,U\right) d^{3}R_{+}d^{3}R_{-}dUda.
\label{twt}
\end{equation}

We note that the distribution of wormholes (\ref{F}) has in general quite
irregular and random behavior and in practical problems it requires some
averaging out $\bar{F}\left( R_{\pm },a,U\right) $, while for a specific
astrophysical object (e.g., a galaxy) it may possess sufficiently strong
fluctuations $\delta F\sim \bar{F}$.

\section{Topological damping of cosmic rays}

In the present section we consider the first terms in the topological
scattering matrix (\ref{tw}) and (\ref{twt}). Those terms define the capture
of particles by wormholes which leads to a specific damping of cosmic rays.
Indeed, let us neglect collisions\footnote{%
For the topological damping the absence of collisions is not essential
though, since they modify merely the function $\widetilde{f}$ in (\ref{ff}).}
and the topological scattering in (\ref{be}) and consider trajectories of
particles $x\left( t\right) =x\left( x_{0},p_{0},t\right) $, $p\left(
t\right) =p\left( x_{0},p_{0},t\right) $. Then we can take variables $\left(
x_{0},p_{0},t\right) $ as new coordinates (instead of $\left( x,p,t\right) $%
) and the equation (\ref{be}) transforms to%
\begin{equation}
\frac{df}{dt}=\alpha \left( r\left( t\right) ,p\left( t\right) ,t\right)
-\left\vert v\left( t\right) \right\vert \beta _{1}\left( r\left( t\right)
\right) f+\left\vert v\left( t\right) \right\vert \int \beta _{2}(\Gamma
,\Gamma ^{\prime })f(\Gamma ^{\prime })d\Gamma ^{\prime },  \label{be2}
\end{equation}%
where $\beta _{1}$ describes the capture of particles, while $\beta _{2}$
describes the remission of the same particles by wormholes. Now if we
consider the case when the source $\alpha \left( t\right) $ produces a
single thin ray and assume that wormholes have isotopic distribution around
the source, then almost all particles captured by wormholes leave the ray
and will radiate from another regions of space and will have different (from
the ray) directions. Then in the first order we can neglect the last term in
r.h.s. of (\ref{be2}) and find the solution in the form%
\begin{equation}
f=e^{-\tau }\widetilde{f},  \label{ff}
\end{equation}%
where $\widetilde{f}$ obey the standard kinetic equation with topological
terms omitted (i.e., $d\widetilde{f}/dt$ $=$ $\partial \widetilde{f}%
/\partial t$ $+$ $\dot{r}\partial \widetilde{f}/\partial r+$ $\dot{p}%
\partial \widetilde{f}/\partial p$ $=$ $\alpha \left( t\right) $), while the
optical depth $\tau \left( t\right) $ describes the damping along the ray%
\begin{equation}
\tau \left( t\right) =\int_{t_{0}}^{t}\beta _{1}\left( r\left( t^{\prime
}\right) \right) \left\vert v\left( t^{\prime }\right) \right\vert
dt^{\prime }=\int_{0}^{\ell }\beta _{1}\left( r\left( s\right) \right) ds,
\end{equation}%
where $\ell $ is the coordinate along the ray.

For astrophysical implications (when the characteristic width of rays $L\gg
a $) we can replace $\delta \left( \xi _{\pm }-a\right) $ in (\ref{tw}) with
$\pi a^{2}\delta \left( \vec{R}_{\pm }-\vec{r}\right) $ (which means that
the absorption of particles occurs at the positions $\vec{R}_{\pm }$, i.e.,
we neglect the throat size $a$). Then, from (\ref{twt}) we find
\begin{equation}
\beta _{1}\left( r\right) =\pi \sum_{n,s=\pm }a_{n}^{2}\delta \left( \vec{R}%
_{s}^{n}-\vec{r}\right) =\pi \int a^{2}n(r,a)da,  \label{dmp1}
\end{equation}%
where $n=n_{+}+n_{-}$, and $n_{s}(r,a)$ $=$ $\int \delta \left( \vec{R}_{s}-%
\vec{r}\right) F\left( R_{\pm },a,U\right) d^{3}R_{+}d^{3}R_{-}dU$. For the
sake of simplicity we consider the case when the distribution of wormholes
reduces to $\overline{F}\left( R_{\pm },a,U\right) $ $=$ $g\left( a\right)
F\left( R_{\pm },U\right) $. Then the value $\beta _{1}\left( r\right) $ can
be expressed via the density of wormholes as
\begin{equation}
\beta _{1}\left( r\right) =\pi \overline{a^{2}}n\left( r\right) ,
\label{dmp}
\end{equation}%
where $\overline{a^{2}}=\int a^{2}g\left( a\right) da$ and $n\left( r\right)
=n_{+}\left( r\right) +n_{-}\left( r\right) $ is the total density of
wormholes $n_{\pm }\left( r\right) =\sum_{n}\delta \left( \vec{R}_{\pm }^{n}-%
\vec{r}\right) $.

\section{Topological bias of a point source}

Consider now the case of a stationary point-like source which radiates
particles in an isotropic way, i.e., $\alpha \left( r,p,t\right) =\lambda
\left( \varepsilon \right) \delta \left( \vec{r}-\vec{r}_{0}\right) $, where
$\varepsilon =\sqrt{p^{2}+m^{2}}$ and $\lambda \left( \varepsilon \right) $
is the distribution of the rate of emission of particles over the momenta.
Then if we neglect the external force ($\dot{p}=0$), collisions, and the
scattering on the wormholes the stationary solution to (\ref{be}) is
\begin{equation}
f_{0}\left( r,p\right) =\frac{m\lambda \left( \varepsilon \right) }{%
p\left\vert r-r_{0}\right\vert ^{2}}\delta \left( \cos \theta -\cos \theta
^{\prime }\right) \delta \left( \varphi -\varphi ^{\prime }\right)  \label{f}
\end{equation}%
where $\theta $, $\varphi $ define the direction of the vector $(\vec{r}-%
\vec{r}_{0})$ and $\theta ^{\prime }$, $\varphi ^{\prime }$ that of $\vec{p}$%
.

When the density of wormholes is low enough the topological term can be
accounted for in the next order which defines the topological bias of the
source $\alpha \rightarrow \alpha +\delta \alpha _{halo}$, where the halo
density is given by%
\[
\delta \alpha _{halo}\left( \Gamma \right) =\left\vert v\right\vert \int
\beta ^{tot}(\Gamma ,\Gamma ^{\prime })f_{0}\left( \Gamma ^{\prime }\right)
d\Gamma ^{\prime }.
\]%
Such a halo has the two terms $\delta \alpha _{halo}\left( \Gamma \right)
=\delta \alpha _{1,halo}+\delta \alpha _{2,halo}$, where the first term
describes the damping (\ref{dmp}) and the second term defines the remission
of particles. The exact form of the halo can be found by the image method as
in Ref. \cite{KS07}. Indeed, if we continue the solution to the whole space
(we recall that the inner region of wormholes $\left\vert \vec{r}-\vec{R}%
_{\pm }^{n}\right\vert <a_{n}$ represents the non-physical region of space),
the wormholes will produce secondary sources of particles. Thus, when we
neglect the throat size ($a\ll R_{\pm }$) and assume the isotropic
distribution over the matrix $U$, then upon averaging over $U$ every
wormhole will radiate in \ the isotropic way which defines the halo as
\[
\delta \bar{\alpha}_{2,halo}\left( \vec{r},\vec{p}\right) =\lambda \left(
\varepsilon \right) B_{2}\left( \vec{r}\right) ,
\]%
where%
\begin{equation}
B_{2}\left( \vec{r}\right) =\sum_{n,s=\pm }\frac{\pi a_{n}^{2}}{\left\vert
\vec{R}_{s}^{n}-\vec{r}_{0}\right\vert ^{2}}\delta \left( \vec{r}-\vec{R}%
_{-s}^{n}\right) ,  \label{b}
\end{equation}%
which defines an additional distribution of particles in the form $f\left(
r,p\right) =f_{0}\left( r,p\right) +\delta \bar{f}\left( r,p\right) $%
\[
\delta \bar{f}\left( r,p\right) =\frac{m\lambda \left( \varepsilon \right) }{%
p}\sum_{n,s=\pm }\frac{\pi a_{n}^{2}}{\left\vert \vec{R}_{s}^{n}-\vec{r}%
_{0}\right\vert ^{2}\left\vert \vec{r}-\vec{R}_{-s}^{n}\right\vert ^{2}}%
\delta \left( \cos \theta _{-s}^{n}-\cos \theta ^{\prime }\right) \delta
\left( \varphi _{-s}^{n}-\varphi ^{\prime }\right) .
\]%
The above expressions can be re-written via the distribution (\ref{F}), e.g.,%
\begin{equation}
B_{2}\left( \vec{r}\right) =\int \frac{\pi a^{2}}{\left\vert \vec{R}-\vec{r}%
_{0}\right\vert ^{2}}N\left( r,R,a\right) d^{3}Rda,  \label{bb}
\end{equation}%
where $N\left( r,R,a\right) =N_{+}+N_{-}$ and $N_{s}$ $=$ $\int \delta
\left( \vec{R}-\vec{R}_{-s}\right) $ $\delta \left( \vec{r}-\vec{R}%
_{s}\right) $ $F\left( R_{\pm },a,U\right) $ $d^{3}R_{+}$ $d^{3}R_{-}$ $dU$
(we point out to the obvious relation $n(r,a)$ $=\int N\left( r,R,a\right)
d^{3}R$ with the distribution $n(r,a)$ in (\ref{dmp1})).

In this manner we see that both functions the damping of cosmic rays (\ref%
{dmp1}) and the distribution of secondary sources (the halo density) (\ref%
{bb}) are determined via the same function $N\left( r,R,a\right) $, i.e.,
the distribution of wormholes which has an irregular (random) behavior.
Together with $N\left( r,R,a\right) $ functions $\beta _{1}\left( r\right) $
and $B_{2}\left( r\right) $ acquire the random character. However, due to
the functional dependence on the only random function $N\left( r,R,a\right) $
such quantities should exhibit a rather strong correlation.

We point out that the interpretatin of the cosmic rays damping possesses an
ambiguity. For instance a suppression of the cosmic ray flux could be also
due to other effects, like multiple scattering in the source itself (e.g.
see Ref. \cite{dfl06} and references therein). Such effects produce
analogous correlation between the damping and the halo of the secondary
sources. Moreover, the halo of the secondary sources (\ref{bb}) is rather
difficult to observe; for the brightness of such a halo is very low (the
intensities of the secondary sources are strongly suppressed by the factor $%
a^{2}/R^{2}$, where $a$ is the effective section of the scatterer and $R$ is
the distance to the scatterer). However, the key point which allows to
disentangle this specific topological damping from other effects is that the
same distribution of wormholes defines the distribution of dark matter which
we discuss in the next section.

\section{Dark matter halos}

As it was demonstrated in Ref. \cite{KS07} (see also discussions in Refs.
\cite{K06,KT07}) the distribution of wormholes (\ref{F}) defines the density
of Dark Matter halos in galaxies as well which is much more easier to
observe. Indeed, in the presence of the gas of wormholes the modification of
the Newton's potential was shown to be accounted for by the topological bias
of sources, i.e., $\delta \left( r-r_{0}\right) \rightarrow $ $\delta \left(
r-r_{0}\right) +b\left( r,r_{0}\right) $, where the halo density $b\left(
r,r_{0}\right) $ is determined via the same distribution of wormholes (\ref%
{F}) by expressions analogous to (\ref{b}), e.g., see for details Ref. \cite%
{KS07}. The form of the bias function $b\left( r,r_{0}\right) $ however
admits the direct measurement by observing rotation curves of galaxies
(e.g., see Refs. \cite{Core} and for the exact form of the bias see Refs.
\cite{K06,KT06}). Indeed, in galaxies the topological bias relates the
densities of dark and luminous matter as

\begin{equation}
\rho _{DM}(r)=\int \bar{b}(r-r^{\prime })\rho _{LM}(r^{\prime
})d^{3}r^{\prime },  \label{rlt}
\end{equation}%
which for the Fourier transforms takes the form $\rho _{DM}(k)$ $=$ $\bar{b}%
(k)\rho _{LM}(k)$. And for a point mass it defines the scale-dependent
renormalization of the dynamic (or the total) mass within the radius $R$ as
\begin{equation}
M_{tot}\left( R\right) /M=1+4\pi \int_{0}^{R}b\left( r\right) r^{2}dr.
\label{M}
\end{equation}

In observations the amount of DM is defined by the mass-to-luminosity ratio $%
M/L$. It is assumed that the luminosity traces the distribution of baryons $%
\rho _{LM}$ which is measured by the observing the surface brightness. E.g.,
spirals can be modeled by an infinitely thin disk with surface mass density
distribution (surface brightness) $\rho _{LM}=\sigma e^{-r/R_{D}}\delta
\left( z\right) $, where $R_{D}$ is the disc radius (the optical radius is $%
R_{opt}=3.2R_{D}$). The total dynamic mass is then defined by the rotation
curve analysis (or by the dispersion of velocities in ellipticals) \cite%
{Core}.

Observations show that the mass-to-luminosity ratio $M_{tot}(r)/L(r)$ for
the sphere of the radius $r$ increases with the distance $r$ from the center
of the galaxy in all galaxies. However if in HSB (high surface brightness)
galaxies this ratio exceeds slightly the unity within the optical disk $%
M(R_{opt})/L\gtrsim 1$ which means that there is a small amount of DM, in
LSB (low surface brightness) galaxies such a ratio may reach $%
M(R_{opt})/L\sim 10^{3}$. Such a correlation between the surface brightness
and the amount of DM in galaxies could give an indirect evidence for the
topological nature of DM; for in accordance to (\ref{dmp}) the amount of
wormholes defines the damping of cosmic rays and analogously the amount of
wormholes defines the amount of dark matter in galaxies \cite{KS07}.
However, the basic mechanism which forms such a feature is different (e.g.,
see Ref. \cite{SL} and references therein). Indeed in smaller galaxies
supernovae are more efficient in removing the gas from the central (stellar
forming) region of a galaxy than in bigger galaxies and this creates the
fact that in smaller objects the disc has a smaller baryonic density (a
lower surface brightness).

In the general case the relationship between the distribution of dark matter
and that of wormholes is rather complicated, e.g., see Ref. \cite{KS07}.
Nevertheless, the renormalization of the intensity a point-like source (\ref%
{M}) allows us to find a rather simple relation between the bias and the
density of wormholes on scales $R\gg \bar{d}$ (where $d=\left\vert \vec{R}%
_{+}-\vec{R}_{-}\right\vert $). We stress that the consideration below has a
rather illustrative (or qualitative) character, while for actual
measurements one has to use the exact relations in Ref. \cite{KS07}.

Indeed, the basic effect of a non-trivial topology is that it cuts some
portion of the volume of the coordinate space. Therefore, the volume of the
physically admissible region becomes smaller, while the density of particles
emitted becomes higher. From the standard flat space standpoint this
effectively looks as if the amplitude of a source renormalizes (\ref{M}).
Consider a ball of the radius $r$ around a point-like source. E.g., for an
isotropic source the number of particles emitted in the unit time in the
solid angle $d\Omega $ $=$ $r^{2}d\phi d\cos \theta $ remains constant $dN$ $%
\sim $ $f_{0}d\Omega $ $=$ $const$, which gives the standard distribution (%
\ref{f}), i.e., $f_{0}\sim I/4\pi r^{2}$.

Let us assume that wormholes have an isotropic distribution around the
source and for the sake of illustration we shall assume that the
distribution has also the structure $\overline{F}\left( R_{\pm },a,U\right) $
$=$ $g\left( a\right) F\left( R_{\pm },U\right) $. Then in the presence of
wormholes the physical volume is
\[
V_{ph}\left( r\right) =\frac{4}{3}\pi \left( r^{3}-\Omega \left( r\right)
\right) ,
\]%
where $\Omega \left( r\right) =4\pi \int a^{3}\int_{0}^{r}n\left( \widetilde{%
r},a\right) \widetilde{r}^{2}d\widetilde{r}da$ defines the portion of the
coordinate volume occupied by wormholes within the the radius $r$ and the
density of wormholes $n\left( r,a\right) $ is defined in (\ref{dmp1}).
Therefore, the actual value of the surface which restricts the ball is $%
S_{ph}\left( r\right) $ $=$ $\frac{d}{dr}V_{ph}\left( r\right) $ and we find
for the density of particles $f$ $\sim I/S_{ph}\left( r\right) $ which
defines the renormalization of the source (\ref{b}) $I\left( r\right)
/I=4\pi r^{2}/S_{ph}\left( r\right) $. Absolutely analogously we can use the
Gauss divergency theorem to estimate the renormalization of the gravity
source. Indeed, the Gauss theorem states that%
\[
\int\limits_{S\left( R\right) }n\nabla GdS=4\pi \int_{r<R}M\delta \left(
r\right) dV=4\pi M,
\]%
where $G$ is the true Green function (or the actual Newton's potential).
Then for isotropic distribution of wormholes it defines the normal
projection of the force as $F_{n}\left( R\right) =n\nabla G=4\pi
M/S_{ph}\left( R\right) $. This can be rewritten as in the ordinary flat
space (in terms of the standard Green function $G_{0}=-1/r$ (i.e., the
standard Newton's law) and the coordinate surface $S_{coor}=4\pi R^{2}$) $%
F_{n}\left( R\right) =M^{\prime }\left( R\right) /R^{2}$, where $M^{\prime
}\left( R\right) /M=4\pi R^{2}/S_{ph}\left( R\right) $ which defines the
bias function in the form (\ref{M}) or%
\begin{equation}
b\left( r\right) =\frac{1}{r^{2}}\frac{d}{dr}\frac{r^{2}}{\frac{d}{dr}%
V_{ph}\left( r\right) }.  \label{bias}
\end{equation}%
We stress again that this function admits the direct measurement in galaxies
\cite{Core,KT06}. Now by make use of the above expression for $V_{ph}\left(
r\right) $ we find the behavior of the dynamic mass for a point source as%
\begin{equation}
\frac{M_{tot}\left( r\right) }{M}=1+\frac{\gamma \left( r\right) }{\left(
1-\gamma \left( r\right) \right) }  \label{MM}
\end{equation}%
where $\gamma (r)=\frac{4}{3}\pi \int a^{3}n\left( r,a\right) da$ which can
be estimated as $\gamma (r)\sim \frac{4}{3}\frac{\overline{a^{3}}}{\overline{%
a^{2}}}\beta _{1}\left( r\right) $. Thus, we see that both quantities the
damping (i.e., the optical depth $\tau $) and the amount of DM (the bias $b$%
) are expressed via the same function $n\left( r,a\right) $.

\section{Conclusions}

For a homogeneous density of wormholes $n\left( r,a\right) =\bar{n}(a)$ and $%
\beta _{1}\left( r\right) =\bar{\beta}_{1}=$ $const$, the damping is
determined merely as $\tau \left( \ell \right) =\bar{\beta}_{1}\ell $ where $%
\ell $ is the coordinate along the ray. Thus, the damping defines the
characteristic scale\footnote{%
We point out that such scale has only statistical meaning, since the actual
distribution of wormholes cannot be utterly homogeneous, otherwise rays
could not reach a sufficiently remote observer. In particular, there is
evidence for the fractal structure of space (e.g., see discussions in Refs.
\cite{K06,KT06} ) which means that there always exist geodesics along which
light propagates almost without the scattering.} $L$ $=$ $1/\bar{\beta}_{1}$~%
$\ $which has the order $L\sim \bar{a}/\gamma $ (where $\gamma =\left( \bar{a%
}/\lambda \right) ^{3}$, $\lambda ^{3}\sim 1/\bar{n}$ is the volume per one
wormhole, and $\bar{a}$ is a characteristic size of throats). The parameter $%
\gamma $ can be extracted from observations of DM in galaxies, while the
scale $\bar{a}$ represents here a free parameter which should be fixed from
some additional and independent considerations. E.g., for the homogeneous
distribution of wormholes the value of $\bar{a}$ defines the amount of dark
energy. Indeed, consider one wormhole in the Minkowski space. Then the
metric can be taken in the form (e.g., see Ref. \cite{KS07})
\[
ds^{2}=dt^{2}-f^{2}\left( r\right) (dr^{2}+r^{2}\sin ^{2}\vartheta d\phi
^{2}+r^{2}d\vartheta ^{2}),
\]%
where $f\left( r\right) =1+\theta (a-r)(\frac{a^{2}}{r^{2}}-1)$ and $\theta
(x)$ is the step function\footnote{%
One can relace $f\left( r\right) $ with a smooth function, this however will
not change the subsequent estimates.}. Both regions $r>a$ and $r<a$
represent portions of the ordinary flat Minkowski space and therefore the
curvature is $R_{i}^{k}\equiv 0$. However on the boundary $r=a$ it has the
singularity which defines the scalar curvature as $R=-T=\frac{1}{a}\delta
\left( r-a\right) $ where $T$ stands for the trace of the stress energy
tensor which one has to add to the Einstein equations to support such a
wormhole. It is clear that such a source violates the averaged null energy
condition, i.e., $T=\varepsilon +3p<0$ (e.g., for the Friedmann space this
results in an acceleration of the scale factor $\sim t^{\alpha }$ with $%
\alpha =\frac{2\varepsilon }{3(\varepsilon +p)}>1$), i.e., represents a form
of dark energy. Every wormhole gives contribution $\int Tr^{2}dr$ $\sim $ $a
$ to the dark energy, while the DE density is $\epsilon $ $=$ $\int an(a,r)da
$ $\sim $ $\gamma /\bar{a}^{2}$. Thus, the parameter $\bar{a}$ can in
principle be extracted from DE density observations. We note however that
one has to be careful in using such a parameter in galaxies, since in the
general case the value $\bar{a}$ is scale dependent (e.g., for the fractal
distribution of wormholes the mean value is unstable).

At the optical radius $r_{opt}$ a galaxy can be considered already as a
point-like source of gravity and, therefore, for estimates we can use (\ref%
{MM}) instead of (\ref{rlt}) and (\ref{bias}). In HSB galaxies the amount of
dark matter within the optical radius $r_{opt}$ is rather small $M/L$ $%
\gtrsim $ $1$ which gives $M_{dyn}\left( r_{opt}\right) /M$ $\sim $ $%
1+\gamma \left( r_{opt}\right) $ with $\gamma \left( r_{opt}\right) \ll 1$
(i.e., $\lambda ^{3}/\overline{a^{3}}\gg 1$) and we can expect the
topological damping to be negligible ($\tau \ll 1$). In LSB galaxies the
mas-to-luminosity ratio may reach $M/L$ $\sim $ $10^{3}$ which gives $\gamma
\left( r_{opt}\right) \sim 1$ and we can expect a considerable damping $\tau
\sim 1$. From the qualitative standpoint this feature agrees with the
observed correlation between the surface brightness and the amount of dark
matter in galaxies which can be considered as an indirect evidence for the
topological nature of Dark Matter. The interpretation of such a feature is
ambiguous though (e.g., see Ref. \cite{SL} and for other effects which lead
to suppression of the cosmic ray flux see Ref. \cite{dfl06}). However we
point out that both quantities $\gamma $ and $\tau $ are functions of the
same random distribution of wormhole $n(a,r)$ and therefore they should
exhibit a rather strong correlation which may allow to verify the
topological nature of DM.

Presumably, astrophysical objects which may also be used to test the
topological nature of DM are large scale extragalactic relativistic jets in
quasars e.g., see Ref. \cite{jets} and references therein. The smaller jets
which are widely observed in active galactic nuclei can also be used in LSB
galaxies, where the amount of DM is considerable. However the crucial step
here is the exact knowledge of the launching mechanism which may allow to
find the discrepancy between the predicted profile of a jet and the actually
observed one.

\section{Acknowledgment}

We acknowledge the advice of referees which helped us to essentially improve
the presentation of this work. For A.A. K this research was supported in
part by the joint Russian-Israeli grant 06-01-72023.

\end{document}